\documentclass[11pt]{amsart}
\usepackage{amsmath}
\usepackage{amssymb}
\usepackage{amscd}
\usepackage{epsfig}
\usepackage{amsxtra}

\newcommand{\bee}{\begin{equation}}
\newcommand{\eee}{\end{equation}}

\newcommand{\be}{\begin{eqnarray}}
\newcommand{\ee}{\end{eqnarray}}
\newcommand{\supp}{\mbox{\rm supp}}

\newcommand{\dens}{\mbox{\rm dens}}

\newcommand{\Vol}{\mbox{\rm Vol}}

\newcommand{\R}{{\mathbb R}}

\newcommand{\Z}{{\mathbb Z}}
\newcommand{\C}{{\mathbb C}}
\newcommand{\XX}{{\mathbb X}}

\newcommand{\mB}{\mathcal B}

  \newtheorem{theorem}{Theorem}[section]
  \newtheorem{lemma}[theorem]{Lemma}
  \newtheorem{prop}[theorem]{Proposition}
  \newtheorem{cor}[theorem]{Corollary}
  \newtheorem{fact}[theorem]{Fact}

  \newtheorem{defi}[theorem]{Definition}
  
  \newtheorem{remark}[theorem]{Remark}

\begin{document}
\title{On the Bragg Diffraction Spectra of a Meyer Set}
\author{Nicolae Strungaru}
\address{Department of Mathematical Sciences, Grant MacEwan University
\\
10700 – 104 Avenue, Edmonton, AB, T5J 4S2;\\
and \\
Institute of Mathematics ``Simon Stoilow'' \\
Bucharest, Romania}
\email{strungarun@macewan.ca}
 \maketitle

\begin{abstract}
Meyer sets have a relatively dense set of Bragg peaks and
for this reason they may be considered as basic mathematical
examples of (aperiodic) crystals. In this paper we investigate the
pure point part of the diffraction of Meyer sets in more detail.
The results are of two kinds. First we show that given a Meyer set
and any intensity $a$ less than the maximum intensity of its Bragg
peaks, the set of Bragg peaks whose intensity exceeds $a$ is
itself a Meyer set (in the Fourier space). Second we show that if a
Meyer set is modified by addition and removal of points in such a
way that its density is not altered too much (the allowable amount
being given explicitly as a proportion of the original density)
then the newly obtained set still has a relatively dense set of  Bragg
peaks.
\end{abstract}
\,

\,

\section{ Introduction }

In 1984 Shechtman, Blech, Gratias, and Cahn announced the discovery of a clear diffraction pattern with a fivefold symmetry, which is impossible in a fully periodic crystal \cite{SCH}, a discovery for which Shechtman was awarded the Nobel prize in Chemistry in 2011.

In the past 25 years hundreds of physical materials with pure point diffraction and no translational symmetry have been found. In 1992 the International Union of Crystallography changed the definition of crystal to "any solid having an essentially discrete diffraction diagram".

Given a point set $\Lambda \subset \R^d$, its diffraction pattern is the measure $\widehat{\gamma}$, where $\gamma$ is the autocorrelation measure of $\Lambda$ (see Section \ref{S3TBMWMSS} for a precise definition of $\gamma$). As any measure, $\widehat{\gamma}$ can be decomposed into its pure point, absolutely continuous and singularly continuous components:

$$ \widehat{\gamma} = (\widehat{\gamma})_{pp} + (\widehat{\gamma})_{ac}+ (\widehat{\gamma})_{sc} \,.$$

The pure point component of the diffraction can be described as $(\widehat{\gamma})_{pp}=\sum_{\chi \in \mB} a_\lambda \delta_\chi$, where $a_\chi >0$ an $\mB$ is the set of {\bf Bragg peaks}. It is usually understood that the diffraction of $\Lambda$ is essentially discrete if $\mB$ is a relatively dense subset of $\R^d$.

A Meyer set is a Delone subset $\Lambda$ of $\R^d$ so that the set of difference vectors $\Delta:= \Lambda- \Lambda$ is uniformly discrete. While this definition looks simple, the full characterization of Meyer sets \cite{MOO} shows that this is actually a strong requirement. Any Meyer set actually has a strong internal order and one would expect this to show in a diffraction experiment. Indeed we proved that the diffraction of any Meyer set shows a relatively dense set of Bragg peaks \cite{NS}. The family of Meyer sets is so far the largest known family of point sets which is both easy to characterize and shows an essentially discrete diffraction.

The goal of this paper is to look closer at the pure point component of the diffraction of a Meyer set $\Lambda$.

In any physical diffraction experiment we cannot see the Bragg peaks of arbitrary small intensities. There is actually a threshold $a>0$ so that we can only see the Bragg peaks of intensity at least $a$. We will call these the $a$-visible Bragg peaks. Formally, the set of $a$-visible Bragg peaks is defined as

$$ I(a):= \{ \chi \in \widehat{\R^d} | \widehat{\gamma}(\{ \chi \}) \geq a \} \,,$$

The main result we prove in this paper is that for a Meyer set $\Lambda$ with autocorrelation measure $\gamma$, and for all $0 < a < \widehat{\gamma}(\{ 0 \})$, the set $I(a)$ of $a$-visible Bragg peaks is a Meyer set.

We also prove on the way that the pure point part $\widehat{\gamma}_{pp}$ is an almost periodic measure in a suitable topology, called the sup topology ( see Definition \ref{suptop} and Definition \ref{supnorm} for the exact definition).

While we are usually interested in the Bragg peaks of high intensity, many point sets exhibit Bragg peaks with intensity as small as possible.
We will show that aperiodic Meyer sets exhibit a much stronger property: If $\Lambda$ is a Meyer set which is not a subset of a periodic crystal, then for any open interval $I \subset [0, \widehat{\gamma}(\{ 0 \})]$, the set of Bragg peaks with intensity in $I$ is a Meyer set.

All these results are collected in the main Theorem in this paper:

{\bf Theorem \ref{aperMS}} Let $\Lambda$ be a Meyer set in $\R^d$, let $\gamma$ be an autocorrelation of $\Lambda$ and let $\Delta := \Lambda- \Lambda$. Then:
\begin{itemize}
\item[i)] Let $0 < \epsilon < 1$. Then, for all $\chi \in \Delta^\epsilon$ we have
$$\widehat{\gamma}(\{ \chi \} \geq (1- \epsilon) \widehat{\gamma}(\{ 0 \}) \,.$$

In particular $\Delta^\epsilon \subset \mB$.

\item[ii)] For each $0 < a < \widehat{\gamma}(\{ 0 \})$ there exists an $\epsilon>0$ and a finite set $F$ so that
$$\Delta^\epsilon \subset I_a(\gamma) \subset \Delta^\epsilon +F \,.$$
\item[iii)]  For each $0 < a < \widehat{\gamma}(\{ 0 \})$, the set $I(a)$ is a Meyer set.
\item[iv)] $\widehat{\gamma}_{pp}$ is a nontrivial sup almost periodic measure.
\item[v)] If $\Lambda$ is not a subset of a fully periodic crystal, then for all $0< b < a < \widehat{\gamma}(\{ 0 \})$ the set
$$\{ \chi | b < \widehat{\gamma}(\{ \chi \}) < a \} \,,$$
is a Meyer set.
\end{itemize}

While we are mainly interested in the diffraction of Meyer sets, we will see in Section \ref{S3TBMWMSS} and Section \ref{S4VBP} that most of the results from Theorem \ref{aperMS} hold for the Fourier Transform $\widehat{\eta}$ of any positive and positive definite measure $\eta$ with Meyer set support and with $\widehat{\eta}_{pp} \neq 0$. Some of the more general results we prove in these sections might be of independent interest to some people.

The second question we study in this paper is what happens if we take a Meyer set $\Lambda$ and change it by removing and adding a "smaller" set. If $\Gamma$ is the set we obtain by this process, then it is easy to see that $\Gamma$ is obtained from $\Lambda$, by removing $\Lambda \backslash \Gamma$, and then adding $\Gamma \backslash \Lambda$. Thus, our change is exactly $\Lambda \bigtriangleup \Gamma$. In Theorem \ref{dens} we prove that that if we start with any Meyer set $\Lambda$, and if the density of our percolation $\Lambda \bigtriangleup \Gamma$ is much smaller than the density of $\Lambda$, then some of the Bragg peaks of $\Lambda$ still show in the diffraction of $\Gamma$.

This paper is organized as follows: In section \ref{S3TBMWMSS} we study the connection between the $\epsilon$-dual characters of a set $\Delta$ and the set of sup almost periods of $\widehat{\eta}_{pp}$, where $\eta$ is a positive definite measure supported inside $\Delta$. In Theorem \ref{epsilonC} we prove that $\Delta^{\epsilon}$ is a subset of the set of $C\epsilon$ sup almost periods of $\widehat{\eta}_{pp}$, for some $C$.
In Section \ref{S4VBP} we use this result to study the connection between $\Delta^{\epsilon}$ and the sets $I_a(\eta)$, while in Section \ref{DMS} we look to the diffraction of Meyer sets. In Section \ref{Cord} we see that under the Meyer assumption, we get a stronger version of the Cordoba result on diffraction of crystals. We conclude the paper by looking to small percolation of Meyer sets.

\section{Preliminaries}\label{S1}

The setting of this paper is $\R^d$, whose dual group is also $\R^d$. To avoid confusion, we will always use $x,y,z$ to denote elements in $\R^d$, and use $\chi, \phi$ and $\psi$ to denote elements in the dual group $\widehat{\R^d} \simeq \R^d$.

The duality between $\R^d$ and $\R^d \simeq \widehat{\R^d}$ is given by

$$< \chi, x >  = e^{-2 \pi i x \cdot \chi }=: \chi(x) \,,$$

where $x \cdot \chi$ denotes the dot product in $\R^d$.

Meyer sets in $\R^d$ were introduced by Y. Meyer in \cite{Meyer}, and fully
characterized by R. V. Moody:

\begin{theorem}{\rm \cite {MOO}} \label{S3CMS} Let $\Lambda \subset \R^d$ be relatively dense. Then the following are equivalent:

\begin{itemize}
\item[i)] $\Lambda$ is a subset of a model set,
\item[ii)] $\Lambda - \Lambda$ is uniformly discrete,
\item[iii)] $\Lambda$ is discrete and there exist a finite set $F$ so that $\Lambda-\Lambda \subset \Lambda+F$,
\item[iv)] For all $0 <\epsilon$, the $\epsilon$-dual set
$$\Lambda^\epsilon:= \{ \chi \in \R^d \mid \left| 1- \chi(x) \right| \leq \epsilon \}$$ is
relatively dense,
 \item[v)] For some $0 < \epsilon < 1/2$,
the $\epsilon$-dual set $\Lambda^\epsilon$ is relatively dense,
\item[vi)] For all $0 <\epsilon$, and any algebraic character $\phi$ on $\R^d$, there exists a continuous character $\chi \in \widehat{\R^d}$ so that

$$\sup_{x \in \Lambda} \left| \chi(x)-\phi(x) \right| \leq \epsilon  \,.$$
\end{itemize}
\end{theorem}

\begin{defi} A set $\Lambda \subset \R^d$ is called a {\bf Meyer set} if $\Lambda$ is relatively dense and  $\Lambda - \Lambda$ is uniformly discrete.
\end{defi}

Any Meyer set verifies all the conditions of Theorem \ref{S3CMS}.

Lets observe that if $\Lambda$ is a Meyer set and $\Delta:= \Lambda- \Lambda$ is the set of relative position vectors in $\Lambda$, then with $F$ given by $iii)$ in Theorem \ref{S3CMS} we have:

$$\Delta- \Delta = (\Lambda- \Lambda) - (\Lambda- \Lambda) \subset (\Lambda+F)- (\Lambda+F)= \Delta +F-F \,.$$

Moreover, $\Delta$  is uniformly discrete  and contains a translate of $\Lambda$, thus it is also relatively dense. Hence

\begin{fact} If $\Lambda$ is a Meyer set and $\Delta= \Lambda- \Lambda$, then $\Delta$ is also a Meyer set.
\end{fact}

In particular, if $\Lambda$ is Meyer, then for all $\epsilon >0$ the set $\Delta^\epsilon$ is relatively dense. This is the key for most results proven in this paper.

Next we will review the mathematics of diffraction.

Let $\omega$ be a translation bounded measure in $\R^d$. Given a van Hove sequence $\{ A_n \}_n$, we define

$$\gamma_n :=\frac{ \omega|_{A_n} * \widetilde{\omega|_{A_n}} }{\Vol(A_n)} \,.$$

It was shown in \cite{BL} that for a translation bounded measure $\omega$, there exists a space ${\mathcal M}_K^C(\R^d)$ which is compacta in the vague topology so that $\gamma_n \in {\mathcal M}_K^C(\R^d)$ for all $n$. It follows that the sequence $\gamma_n$ always has cluster points.

\begin{defi} Any cluster point $\gamma$ of the sequence $\gamma_n$ is called an {\bf autocorrelation} of $\omega$.

The measure $\gamma$ is positive definite, thus Fourier Transformable. Its Fourier Transform $\widehat{\gamma}$ is a positive measure, called the {\bf diffraction measure} of $\omega$.
\end{defi}

If $\gamma$ is an autocorrelation of $\omega$, by eventually replacing $\{ A_n \}$ by a subsequence we can always assume that

$$\gamma = \lim_n \gamma_n \,.$$

Different choices of van Hove sequences could lead to different autocorrelation measures, and thus also to different diffraction measures. Anyhow, for the results we prove in this paper, the choice of autocorrelation will be irrelevant, if one picks a different autocorrelation the same result will still hold.

If $\supp(\omega) \subset \Lambda$ for a Meyer set $\Lambda$, then for all $n$ we have $\supp(\gamma_n) \subset \Lambda-\Lambda =: \Delta$. Since $\Delta$ is uniformly discrete, it follows that $\supp(\gamma) \subset \Delta$. Thus we get:

\begin{fact} If $\supp(\omega)$ is a subset of a Meyer set $\Lambda$, then the support of any autocorrelation $\gamma$ of $\Lambda$ is a subset of $\Delta:= \Lambda- \Lambda$.
\end{fact}

Given any $\omega$, and an autocorrelation $\gamma$, the diffraction measure $\widehat{\gamma}$ it can be decomposed in the discrete, absolutely continuous and singularly continuous components:

$$\widehat{\gamma}= \widehat{\gamma}_{pp}+\widehat{\gamma}_{ac}+\widehat{\gamma}_{sc} \,.$$

The discrete component $\widehat{\gamma}_{pp}$ can be written as

$$\widehat{\gamma}_{pp}= \sum_{\chi \in \mB} \widehat{\gamma}(\{ \chi \}) \delta_\chi \,,$$

where

$$\mB:= \{ \chi \in \R^d | \widehat{\gamma}(\{ \chi \}) \neq 0 \} \,.$$

$\mB$ is called the set of {\bf Bragg peaks} of $\omega$.

The intensity of Bragg peaks can be calculated by the following formula:

\begin{theorem}\cite{HOF} \cite{LR}\label{HOFave} Let $\gamma$ be a translation bounded Fourier Transformable measure. Then, for all $\chi \in \R^d$ we have
\begin{equation}\label{EQHOF}
\widehat{\gamma}(\{ \chi\}) = \lim_{n \to \infty} \frac { \int_{A_n}
\overline{\chi}(x) d \gamma(x) }{\Vol (A_n )} \,.\end{equation}
\end{theorem}

\section{$\epsilon$-dual characters and sup almost periodicity}\label{S3TBMWMSS}

In this section we show that for a Fourier transformable measure $\eta$ on $\R^d$, with $\supp(\eta) \subset \Delta$, the sets $\Delta^\epsilon$ of $\epsilon$-dual characters are sup almost periods for $\widehat{\eta}_{pp}$.

For the entire section $\{ A_n \}$ is a fixed van Hove sequence.

The main tool we are going to use in this paper is the following result:

\begin{theorem} \label{S31} Let $\eta$ be a Fourier transformable measure and let $ \supp(\eta) \subset \Delta$. Then, there exists a constant $C \geq 0$ such that
 for all $0 < \epsilon, \chi \in \R^d$ and for all
$\psi \in \Delta^{\epsilon}$ we have

\begin{equation}\label{EQS31}
\lvert \widehat{\eta}(\{ \psi+\chi\})-\widehat{\eta}(\{\chi\}) \rvert \leq C \epsilon
\end{equation}

Moreover, if $\eta$ is positive, (\ref{EQS31}) will hold for $C = \widehat{\gamma}(\{ 0 \})$.
\end{theorem}

\noindent {\sc \bf Proof:} Let

$$C= \limsup_{n \to \infty}  \frac { \lvert \eta \rvert (A_n ) }{\Vol (A_n ) } \,.$$

Since $\eta$ is translation bounded, $0 \leq C<   \infty$.

Let $\epsilon >0$. Then, for any $\psi \in
\Delta^{\epsilon}$, any $\chi \in \R^d$ and $x \in \Delta$ we have:

$$\lvert  (\psi+\chi)(x)  - \chi(x)  \rvert=\lvert  \chi(x)\left( \psi(x)  - 1 \right)  \rvert < \epsilon\,.$$

Combining this result with Theorem \ref{HOFave}, and using $\supp(\eta) \subset \Delta$ we get:
\begin{equation}\label{EQAP}
\begin{split} \lvert \widehat{\eta}(\{ \psi+\chi\})
&-\widehat{\eta}(\{\chi\}) \rvert \leq \lim_{n \to \infty} \frac {
\int_{A_n}  \lvert \overline{(\psi+\chi)}(x) -\overline{\chi}(x)
\rvert  \, d  \lvert
\eta  \rvert (x) }{\Vol (A_n ) } \\
&=\lim_{n \to \infty} \frac {
\int_{A_n \cap \Delta }  \lvert \overline{(\psi+\chi)}(x) -\overline{\chi}(x)
\rvert  \, d  \lvert
\eta  \rvert (x) }{\Vol (A_n ) } \\
&\leq \lim_n \frac { \int_{A_n \cap \Delta } \epsilon \, d \lvert \eta \rvert(x)
}{\Vol (A_n ) } \leq C \epsilon \,.
\end{split}
\end{equation}

If $\eta$ is positive, then

$$C= \limsup_{n \to \infty}  \frac { \lvert \eta \rvert (A_n ) }{\Vol (A_n ) }= \lim_{n \to \infty}  \frac { \eta  (A_n ) }{\Vol (A_n ) }= \widehat{\eta}(\{ 0 \}) \,,$$

which proves the last claim.

\qed

\begin{remark} Later in the paper we will need sometimes to assume that the constant $C$ in Theorem \ref{S31} satisfies $C>0$. This can be done, since it is easy to see that we can replace $C$ in Theorem \ref{S31} by any larger number.
\end{remark}

The property we got in Theorem \ref{S31} is very similar to the notion of almost periods. We introduce now a topology on the space of discrete translation bounded measures for which, (\ref{EQAP}) is equivalent to $\psi$ being an $C\epsilon$-almost period.

\begin{defi}\label{supnorm} Let ${\mathcal M}^\infty_{pp}(\R^d)$ denote the space of discrete translation bounded measures. We define a norm on ${\mathcal M}^\infty_{pp}(\R^d)$ by

\begin{equation}\label{EQsup} \| \mu \|_\infty:= \sup_{x \in \R^d} \left| \mu(\{ x \}) \right|
\end{equation}

This is called the {\bf sup norm} on ${\mathcal M}^\infty_{pp}(\R^d)$, and we will refer to the topology defined by this measure as being the {\bf sup topology}.
\end{defi}

One can observe that the definition $\| \cdot \|_\infty$ of \ref{EQsup} makes sense for any measure in ${\mathcal M}^\infty(\R^d)$. Anyhow, on this space $\| \cdot \|_\infty$ is a semi-norm, and it is easy to check that $\| \mu \|_\infty =0$ if and only if $\mu$ is a continuous measure. For this reason we are only interested in the sup norm of a discrete measure.

Now we can define the notion of almost periods for a measure in this topology.

\begin{defi} \label{suptop}
Let $\mu \in {\mathcal M}^\infty_{pp}(\R^d)$. We say that $t \in \R^d$ is an {\bf $\epsilon$-almost period} for $\mu$ if

$$ \| \mu -T_t \mu \|_\infty < \epsilon \,,$$

where $T_t$ denotes the translate by $t$ operator.

We denote by $P^\infty_\epsilon(\mu)$ the set of $\epsilon$-almost periods of $\mu$, that is

$$P^\infty_\epsilon(\mu):= \{ t \in\R^d | \|\mu -T_t \mu \|_\infty < \epsilon \} \,.$$

The measure  $\mu \in {\mathcal M}^\infty_{pp}(\R^d)$ is called {\bf sup almost periodic} if for all $\epsilon >0$ the sets $P^\infty_\epsilon(\mu)$ are relatively dense.

\end{defi}

An immediate consequence of Theorem \ref{S31} is:

\begin{theorem}\label{epsilonC} Let $\eta$ be any Fourier transformable measure and $\Delta$ be so that $\sup(\eta) \subset \Delta$. Then
\begin{itemize}
\item[i)] There exists a $C>0$ so that , for all $\epsilon >0$ we have
$$\Delta^\epsilon \subset P_{C \epsilon}^\infty ( \widehat{\eta}_{pp}) \,.$$
 \item[ii)] If $\Delta$ is a Meyer set, then $ \widehat{\eta}_{pp}$ is sup almost periodic.
\end{itemize}
\end{theorem}
\noindent {\sc \bf Proof:}

$i)$ Let $C_0$ be the constant from Theorem \ref{S31}, and let $C >C_0$ be any number. Then, by Theorem \ref{S31}, for all $\chi \in \Delta^\epsilon$ we have

$$ \| \widehat{\eta}_{pp} -T_\chi \widehat{\gamma}_{pp} \|_\infty \leq C_0 \epsilon < C \epsilon \,.$$

$ii)$: Since $\Delta$ is a Meyer set, all the sets $\Delta^\epsilon$ are relatively dense by theorem \ref{S3CMS}. Thus by $i)$, all the sets $P_{C \epsilon}^\infty ( \widehat{\eta}_{pp})$ are relatively dense.

\qed

\section{Positive Definite Measures with Meyer set support.}\label{S4VBP}

For this entire section $\Delta$ is a Meyer set and $\eta$ is a positive definite measure with $\supp(\eta) \subset \Delta$ and $\widehat{\eta}_{pp} \neq 0$. We will study the connection between the sets $\Delta^\epsilon$, and the set $\mB:= \{ \chi | \widehat{\eta}(\{ \chi \}) \neq 0 \}\,.$

First lets look at the assumptions we make.

The Meyer condition of $\Delta$ is needed since most of our proofs will be based on the relatively denseness of $\Delta^\epsilon$.

The positive definiteness of $\eta$ makes some of the proofs easier. One can probably generalize the results in this section to Fourier Transformable measures, by looking at $| \widehat{\eta
}(\chi\})|$, anyhow the main application we are interested is the case when $\eta$ is an autocorrelation measure. Thus in all cases we are interested, $\eta$ is positive definite, and since this assumption eliminates some complications from the proofs, we will work in this case.

Finally the condition $\widehat{\eta}_{pp}\neq 0$ is needed in order to make sure that $\mB \neq \emptyset$, otherwise there is nothing to say about it. In general this condition is easy to check, as shown in \cite{NS}:

\begin{theorem}\cite{NS} Let $\eta$ be a Fourier Transformable translation bounded measure with $\supp(\eta)-\supp(\eta)$ uniformly discrete. Then $\widehat{\eta}_{pp}=0$ if and only if
$$\lim_{n} \frac{ | \eta | ( A_n) }{ \Vol(A_n)} =0 \,.$$
\end{theorem}

We need to introduce few definitions first :

\begin{defi}
Let $\eta$ be a positive definite measure. We define

$$I_{sup}(\eta) := \sup_{\chi \in \R^d}   \widehat{\eta}(\{ \chi \}) \,.$$
\end{defi}

Note that since $\eta$ is positive definite, the measure $\widehat{\eta}$ is positive and hence $I_{sup}(\eta)=\| \widehat{\eta}_{pp} \|_\infty \,.$ In particular,
$I_{sup}(\eta) \geq 0$, with equality if and only if $\widehat{\eta}_{pp}=0$.
Moreover, if $\eta$ is positive and positive definite, then it follows immediately from Theorem \ref{HOFave} that

$$I_{sup}(\eta)= \widehat{\eta}(\{0\}) \,.$$

Also $\chi \in \mB$ if and only if $\widehat{\eta}(\{ \chi \}) \in (0, I_{sup}(\gamma) ] \,.$

As we said, the goal of this section is to study connection between $\mB$ and $\Delta^\epsilon$. We will do this, by looking to the elements of $\mB$ which are not arbitrary close to $0$:

\begin{defi} Let $a \in \R$. For a positive definite measure $\eta$ we define

$$I_a(\eta):= \{ \chi \in \R^d | \widehat{\eta}(\{ \chi \}) \geq a \} \,.$$
\end{defi}

Lets observe that for $a > I_{sup}(\eta)$ we have $I_a(\eta)= \emptyset$, while for $a \leq 0$ we have $I_a(\eta)= \R^d$. Thus, we are only interested in $I_a(\eta)$ for $0< a \leq I_{sup}(\eta)$.

The main result in this section is:

\begin{theorem}\label{S3P2} Let $\eta$ be a positive definite measure, $\Delta$ a Meyer set with $\supp(\eta) \subset \Delta$. If $\widehat{eta}_{pp} \neq 0$, then there exists a $C>0$ so that:
\begin{itemize}
\item[i)] For any $\epsilon >0$ we have:
$$ I_{a}(\eta) \pm \Delta^{\epsilon} \subset I_{a - C\epsilon}(\eta) \,.$$
\item[ii)] For each $0 < a < I_{sup}( \eta)$, exists an $0< \epsilon(a)$, so that for each $0< \epsilon < \epsilon(a)$ there exists a $\chi \in \R^d$ and a finite set $F$ for which
$$\chi+\Delta^\epsilon \subset I_a(\eta) \subset \Delta^\epsilon +F \,.$$
Moreover, if $\eta$ is positive, $\chi$ can be chosen to be $0$.
\item[iii)] For all $0 < a < I_{sup}( \eta)$, $I_a(\eta)$ is a Meyer set.
\end{itemize}
\end{theorem}
\noindent {\sc \bf Proof:}

Let $C_0$ be the constant given by Theorem \ref{S31}, and let $C>C_0$.

{\bf i)} Follows trivially from Theorem \ref{S31}, since
$\Delta^{\epsilon} =- \Delta^{\epsilon}$.

\bf ii)\rm  We will see that the first inclusion works as long as $2 C \epsilon  < a$, while the second works for all epsilon for which
$2 C \epsilon < \min\{ a, I_{sup}(\eta) -a \}$.

Thus lets set $$\epsilon(a):= \frac{ \min\{ a, I_{sup}(\eta) -a \}}{2C+1} \,.$$

Let $0< \epsilon < \epsilon(a)$. Then $$2 C \epsilon < \min\{ a, I_{sup}(\eta) -a \} \,.$$

By the definition of $I_{sup}(\eta)$, there exists a $\chi$ so that $\widehat{\eta}(\{
\chi \}) >I_{sup}(\eta) - C\epsilon$.

Note that if $\eta$ is positive, we can chose $\chi=0$ since, in this case, $I_{sup}(\eta)=\widehat{\eta}(\{ 0\})$.

{\it Step 1}: We show that $\chi+\Delta^\epsilon \subset I_a(\eta)$:

Let $\psi \in \Delta^\epsilon$ be arbitrary. By Theorem \ref{S31} we have

$$\left| \widehat{\eta}(\{ \psi+\chi \})- \widehat{\eta}(\{ \chi \}) \right| < C \epsilon \,.$$

Since the measure $ \widehat{\eta}$ is positive, by the triangle inequality we get:

$$\widehat{\eta}(\{ \psi+\chi \}) \geq \widehat{\eta}(\{ \chi \}) - \left| \widehat{\eta}(\{ \psi+\chi \})- \widehat{\eta}(\{ \chi \}) \right| \geq \widehat{\eta}(\{
\chi \}) - C\epsilon \,.$$

Using now the fact that $\widehat{\eta}(\{ \chi \}) >I_{sup}(\eta) - C\epsilon$ we get

$$\widehat{\eta}(\{ \psi+\chi \}) \geq  I_{sup}(\eta) - 2 C\epsilon > a \,,$$

which proves that

$$\psi+\chi \in I_a(\eta) \,.$$

{\it Step 2}: We show that there exists a finite set $F$ so that $I_a(\eta) \subset \Delta^\epsilon +F$:

First lets pick $b:= a-C\epsilon$. Then $b>0$.

Since $\Delta^\epsilon$ is relatively dense, there exists a compact set $K$ so that $\Delta^\epsilon + K =\R^d$.

The idea for the remaining of the proof is very simple: Given and element $\psi \in I_a$, the set $\psi- \Delta^\epsilon$ must meet $K$, and at any intersection point $\phi$ we have $\widehat{\eta}(\{ \phi \}) \geq b$. But, by the regularity of $\widehat{\eta}$ we can only have finitely many elements $\phi$ in $K$ so that $\widehat{\eta}(\{ \phi \}) \geq b$, and this will define our finite set $F$.

Let
$$F:= \{ \phi \in K | \widehat{\eta}(\{ \phi \})> b \} = I_b(\eta) \cap K \,.$$ Then $F$ is a finite set.

By $i)$ we know that $I_a(\eta)-\Delta^{\epsilon} \subset I_b(\eta)$.

Let $\phi \in I_a(\eta)$. Since $\phi \in \R^d= \Delta^\epsilon + K$, we can write $\phi=\psi+\tau$ with $\psi \in \Delta^\epsilon$ and $\tau \in K$.

Then we have

$$\tau = \phi- \psi \in I_a(\eta)- \Delta^\epsilon \subset I_b (\eta) \,.$$

Thus, $\tau \in I_b(\eta) \cap K = F$.

We proved that any $\phi \in I_a(\eta)$ can be written as $\phi=\psi+\tau$ with $\psi \in \Delta^\epsilon$ and $\tau \in F$. This proves that

$$ I_a(\eta) \subset \Delta^\epsilon +F \,.$$

$iii)$

Pick some $0 < \epsilon < \min \{ \epsilon(a), \frac{1}{2} \}$.

By $ii)$, we have
$$\chi+\Delta^\epsilon \subset I_a(\eta) \subset \Delta^\epsilon +F \,.$$
for some $\chi \in \R^d$ and some finite set $F$.

Since $\Delta$ is a Meyer set, and $0< \epsilon < \frac{1}{2}$, the set $\Delta^\epsilon$ is also a Meyer set, and thus so is $\Delta^{\epsilon}+F$.

Hence $I_a(\eta)$ is a subset of a Meyer set. Moreover, it is relatively dense, since it contains $\chi + \Delta^\epsilon$.

Therefore $I_a(\eta)$ is a Meyer set.

\qed

For the reminding of the section we will study when $I_a(\eta)=\mB$ for some $a>0$. We will prove that if this happens, then both $\Delta$ and $\mB$ are subsets of finitely many translates of lattices.
More exactly, we will show that if $I_a(\eta)=\mB$ for some $a>0$, then there exists a lattice $L$ and finite sets $F_1 \subset \R^d$ and $F_2 \subset \widehat{\R^d}$ so that

$$\Delta \subset L+ F_1 \,;\, \mB \subset L^\ast+ F_2 \,,$$

where $L^\ast$ is the dual lattice of $L$.

This result will prove that, unless $\eta$ is supported inside a fully periodic set, for each $\epsilon >0$, we can find a $\chi$ so that $0< \widehat{\eta}(\{ \chi \}) < \epsilon$.

We need first to prove a simple Lemma:

\begin{lemma}\label{S24}
Let $\Lambda \subset \Gamma \subset \R^d$ be so that $\Lambda$ is relatively dense
and $\Gamma$ has finite local complexity. Then, there exists a finite
set $F$ so that $\Gamma \subset \Lambda+F$.
\end{lemma}

\noindent {\sc \bf Proof:} Let $K$ be compact so that $\Lambda+K= \R^d$.
Let $F:= (\Gamma-\Gamma) \cap K$. Then $F$ is finite.

We prove that $\Gamma \subset \Lambda+F$.

Let $x \in \Gamma$. Since $x \in \R^d=\Lambda+K$ we can write  $x= y+f$ with $y \in \Lambda$ and
$f \in K$.

Then $f=x-y \in \Gamma-\Lambda \subset \Gamma-\Gamma$. Thus $f \in (\Gamma-\Gamma) \cap K =F$.

Hence

$$x=y+f \,;\, y \in \Lambda , f \in F \,.$$
\qed

We will now show that if $I_a(\eta)=I_b(\eta)$ for some $0 < b <a < I_{sup}(\eta)$  then there exists a lattice $L$ and two finite sets $F_1, F_2$ so that $\Delta \subset L+F_1$ and $I_a(\eta) \subset L^* +F_2$. The idea behind this proof is simple: pick an $\epsilon>0$ so that $I_a(\eta) \pm \Delta^\epsilon \subset I_b(\eta)= I_a(\eta)$. Thus, any $\chi \in \Delta^\epsilon$ is a period for $I_a(\eta)$, and hence so is the group $L^*$ generated by $\Delta^\epsilon$.  It is easy to prove that this group is a lattice, and then Lemma \ref{S24} completes the claim.

\begin{prop} \label{prop123} Let $\eta$ be a positive definite measure and let $\supp(\eta) \subset \Delta$. If $\Delta$ is a Meyer set and $\emptyset \neq I_a(\eta)=I_b(\eta) $ for some $0< b <a$, then there exists a lattice $L$, with dual lattice $L^\ast$, and finite sets $F_1, F_2$ so that

$$\Delta \subset L+F_1 \,;\, I_a(\eta) \subset L^\ast + F_2 \,.$$
\end{prop}

\noindent {\sc \bf Proof:} By Theorem \ref{S3P2}, there exists some $0< \epsilon < \frac{1}{2}$ so that,
$$I_a(\eta)-\Delta^\epsilon \subset I_b(\eta) \,.$$

Since $I_a(\eta) =I_b(\eta)$ we get

$$I_a(\eta)- \Delta^\epsilon \subset I_a(\eta) \,.$$

Let $L^{*}$ be the group generated by $\Delta^\epsilon$. Then

$$I_a(\eta)-L^\ast \subset I_a(\eta) \,.$$

We claim that $L^\ast$ is a lattice in $\R^d$. Since $\Delta^\epsilon$ is relatively dense, we get that $L^\ast$ is a relatively dense subgroup of $\R^d$. To prove that $L^\ast$ is a lattice, we need to show that it is also discrete. But this follows from $$I_a(\eta)-L^\ast \subset I_a(\eta) \,.$$

Indeed, for some $\chi \in I_a(\eta)$ we get that

$$\chi - L^\ast \subset I_a(\eta) \,.$$

But then, by Theorem \ref{S3P2} the set $I_a(\eta)$ is uniformly discrete, and thus, so is $L^\ast$.

We proved so far that $L^\ast$ is a lattice.

Since $\chi - L^\ast \subset I_a(\eta), L^\ast$ is relatively dense and $I_a(\eta)$ is a Meyer set, it follows from  Lemma \ref{S24} that

$$I_a(\eta) \subset L^\ast + F_2 \,,$$

for some finite set $F_2$. This proves the second part of our claim.

Now we prove the rest of the claim. Let $L$ be the dual lattice of $L^\ast$.

Since $\Delta^\epsilon \subset L^\ast$, we get \cite{MOO}

$$(L^\ast)^\epsilon \subset \Delta^{\epsilon\epsilon} \,.$$

But for any $0<\epsilon < \frac{1}{2}$, the $\epsilon$ dual set of a lattice is the dual lattice \cite{MOO}, and thus

$$L \subset \Delta^{\epsilon\epsilon} \,.$$

Using the fact that $L$ is a lattice, and  $\Delta^{\epsilon\epsilon} $ is a Meyer set, we get again by Lemma \ref{S24} that

$$\Delta^{\epsilon\epsilon} \subset L+F_1 \,,$$

for some finite set $F_1$.

Our claim follows now from

$$ \Delta \subset \Delta^{\epsilon\epsilon}  \,.$$

\qed

We conclude the section with an interesting consequence of Proposition \ref{prop123}:

\begin{cor} \label{cor123} Let $\eta$ be a positive definite measure and let $\supp(\eta) \subset \Delta$. If $\Delta$ is a Meyer set and $I_a(\eta)=\mB \neq \emptyset$ for some $a>0$, then there exists a lattice $L$, with dual lattice $L^\ast$, and finite sets $F_1, F_2$ so that

$$\Delta \subset L+F_1 \,;\, \mB \subset L^\ast + F_2 \,.$$
\end{cor}

\section{Diffraction under the Meyer condition}\label{DMS}

If $\gamma$ is the autocorrelation of some translation bounded measure $\mu$, and $\supp(\gamma)$ is a (subset of a) Meyer set, the results proven in Section \ref{S4VBP} yield some interesting consequences about the set of Bragg peaks in the diffraction of $\mu$.

Lets observe that in this case

$$I_a(\gamma)= \{ \chi \in \R^d | \widehat{\gamma}(\{ \chi \}) \geq a \} \,,$$

is exactly the set of Bragg peaks of intensity at least $a$. We will call this set, the set of {\bf $a$ -visible Bragg peaks}.

Lets start by looking to the diffraction of a Meyer set $\Lambda$. If $\gamma$ is an autocorrelation of $\Lambda$, then $\gamma$ is positive and  $\widehat{\gamma}_{pp}$ is nontrivial (see \cite{NS} for example).

\begin{theorem}\label{aperMS} Let $\Lambda$ be a Meyer set in $\R^d$, let $\gamma$ be an autocorrelation of $\Lambda$ and let $\Delta := \Lambda- \Lambda$. Then:
\begin{itemize}
\item[i)] Let $0 < \epsilon < 1$. Then, for all $\chi \in \Delta^\epsilon$ we have
$$\widehat{\gamma}(\{ \chi \} \geq (1- \epsilon) \widehat{\gamma}(\{ 0 \}) \,.$$

In particular $\Delta^\epsilon \subset B$.

\item[ii)] For each $0 < a < \widehat{\gamma}(\{ 0 \})$ there exists an $\epsilon>0$ and a finite set $F$ so that
$$\Delta^\epsilon \subset I_a(\gamma) \subset \Delta^\epsilon +F \,.$$
\item[iii)]  For each $0 < a < \widehat{\gamma}(\{ 0 \})$, the set $I(a)$ is a Meyer set.
\item[iv)] $\widehat{\gamma}_{pp}$ is a nontrivial sup almost periodic measure.
\item[v)] If $\Lambda$ is not a subset of a fully periodic crystal, then for all $0< b < a < \widehat{\gamma}(\{ 0 \})$ the set
$$\{ \chi | b < \widehat{\gamma}(\{ \chi \}) < a \} \,,$$
is a Meyer set.
\end{itemize}
\end{theorem}
\noindent {\sc \bf Proof:}

[i)] By Theorem \ref{S31}, since $\gamma$ is positive we have

$$\left| \widehat{\gamma}(\{ 0 \} ) - \widehat{\gamma}(\{ \chi \}) \right| \leq \epsilon \widehat{\gamma}(\{ 0\}) \,.$$

which implies the desired inequality.

$ii)$ and $iii)$ follow immediately from Theorem \ref{S3P2}.

$iv)$ We know that $\widehat{\gamma}_{pp}$ is nontrivial by \cite{NS}. Hence, this claim is an immediate consequence of Theorem \ref{epsilonC}.

$v)$ Let $c= \frac{a+b}{2}$ and let $0 < \epsilon < \frac{b-a}{4}$. Then $b < c - 2\epsilon < c+2 \epsilon < a$. Note that $c$ is exactly the midpoint of $(b,a)$.

We will prove this claim in two steps. We will first find some $\chi$ so that $\left| \widehat{\gamma}(\{ \chi \}) -c \right| < c$, and then, we will show that for all elements in $\chi+\Delta^{\frac{\epsilon}{C}}$ the intensity is within $2\epsilon$ of $c$, thus between $b$ and $a$.

Since $\Delta$ is not a subset of a fully periodic set, by Proposition \ref{prop123}, we get that $I_{c - \epsilon}(\gamma) \neq I_{c + \epsilon}(\gamma)$. Thus

$$I_{c + \epsilon}(\gamma) \subsetneqq I_{c - \epsilon}(\gamma) \,.$$

Pick some $\psi \in I_{c - \epsilon}(\gamma) \backslash I_{c + \epsilon}(\gamma)$. Then

$$ \left| \widehat{\gamma}(\{ \psi \}) -c \right| < \epsilon \,.$$

Now, for all $\varphi \in \Delta^{\frac{\epsilon}{\widehat{\gamma}(\{ 0\})}}$  we have

$$\left| \widehat{\gamma}(\{ \varphi+\chi \})-c \right| \leq \left| \widehat{\gamma}(\{ \varphi+\chi \})- \widehat{\gamma}(\{ \chi \}) \right| + \left| \widehat{\gamma}(\{ \chi \})- c \right| < \epsilon +\epsilon=2\epsilon \,.$$

Thus

$$\psi + \Delta^{\frac{\epsilon}{\widehat{\gamma}(\{0\})}} \subset \{ \chi | b < \widehat{\gamma}(\{ \chi \}) < a \} \,.$$

Now our claim is immediate. The set  $\{ \chi | b < \widehat{\gamma}(\{ \chi \}) < a \}$ contains the relatively dense set $\psi + \Delta^{\epsilon/C}$, and is a subset of $I_{b}(\gamma)$, which is a Meyer set. Thus, $\{ \chi | b < \widehat{\gamma}(\{ \chi \}) < a \}$ is a  Meyer set \cite{MOO}.
\qed

A Meyer set $\Lambda$ can have multiple autocorrelations. If $\gamma_1$ and $\gamma_2$ are two different autocorrelations of a Meyer set $\Lambda$, then the corresponding sets $I(a)$ and $B$ can be very different, but Theorem \ref{aperMS} can be applied for each of them. The following is an interesting consequence of Theorem \ref{aperMS} ii):

\begin{cor} Let $\Lambda$ be a Meyer set and let $\gamma_1, \gamma_2$ be two autocorrelations of $\Lambda$. Let $0< a_1 < \widehat{\gamma_1}(\{ 0\})$ and $0< a_2 < \widehat{\gamma_2}(\{ 0\})$. Then, there exists a finite set $F$ so that

$$I_{a_1}(\gamma_1) \subset I_{a_2}(\gamma_2)+F \,;\, I_{a_2}(\gamma_2) \subset I_{a_1}(\gamma_1)+F \,.$$
\end{cor}

It is easy to see that most of the arguments we did in the proof of Theorem \ref{aperMS} hold for the larger class of weighted Dirac combs with Meyer set support. The only facts we used in the proof of Theorem \ref{aperMS}, which don't necessarily hold for weighted combs are the positivity of the autocorrelation, and the existence of Bragg peaks in the diffraction. If we put these two conditions as extra requirements, we get:

\begin{theorem} Let $\Lambda$ be a Meyer set in $\R^d$, let $\omega= \sum_{x \in \Lambda} \omega(x) \delta_x$ be a translation bounded measure, let $\gamma$ be an autocorrelation of $\omega$ and let $\Delta := \Lambda- \Lambda$. If $\widehat{\gamma}_{pp}$ is nontrivial then:
\begin{itemize}
\item[i)] There exists an $\epsilon_0 >0$ and a character $\chi$ so that for all $0< \epsilon <\epsilon_0$ we have $\chi+\Delta^\epsilon \subset \mB$.
\item[ii)] For each $0 < a < I_{sup}(\gamma)$ there exists an $\epsilon>0$ and a finite set $F$ and some $\chi$ so that
$$\chi + \Delta^\epsilon \subset I_a(\gamma) \subset \Delta^\epsilon +F \,.$$
\item[iii)]  For each $0 < a < \widehat{\gamma}(\{ 0 \})$, the set $I(a)$ is a Meyer set.
\item[iv)] $\widehat{\gamma}_{pp}$ is a nontrivial sup almost periodic measure.
\item[v)] If $\Lambda$ is not a subset of a fully periodic crystal, then for each $0< b < a < \widehat{\gamma}(\{ 0 \})$ the set
$$\{ \chi | b < \widehat{\gamma}(\{ \chi \}) < a \} \,,$$
is a Meyer set.
\end{itemize}
\end{theorem}

\noindent {\sc \bf Proof:} We only need prove $i)$, everything else is obvious.

Since $\widehat{\gamma}_{pp}$ is not trivial, we can find some $\chi$ so that

$$ \widehat{\eta}(\{ \chi \}) > 0 \,.$$

By Theorem \ref{S31}, there exist a $C$ do that for all $\epsilon >0 $, all $\psi \in \Delta^\epsilon$ and $\chi \in \R^d$ we have

$$\left| \widehat{\eta}(\{ \psi+\chi \})- \widehat{\eta}(\{ \chi \}) \right| \leq C\epsilon  \,.$$

Exactly like in the proof of Theorem \ref{aperMS}, if we make $C \epsilon <  \widehat{\eta}(\{ \chi \}) $ we get $\widehat{\eta}(\{ \psi+\chi \}) \neq 0$. Thus, picking any $0 < \epsilon_0$ so that $C \epsilon_0 <  \widehat{\eta}(\{ \chi \}) $ completes the proof.
\qed

The main requirement for most proofs is not that the point set $\Lambda$ or the measure $\omega$ is supported on a Meyer set. Instead we only need the autocorrelation $\gamma$ to have a Meyer set support, which is a weaker requirement. This can happen without the original measure having a Meyer set support. A simple such example is

$$\Lambda:= \{ n +\frac{1}{n} | n \in \Z \backslash \{ 0 \} \} \,.$$

This $\Lambda$ is a non-Meyer Delone set, which has an unique autocorrelation $\gamma=\delta_\Z$.

In this situation we can still prove the following:

\begin{theorem} Let $\omega$ be a translation bounded measure, let $\gamma$ be an autocorrelation of $\omega$. If $\supp(\gamma) \subset \Delta$ for some Meyer set $\Delta$ and if $\widehat{\gamma}_{pp}$ is nontrivial then:
\begin{itemize}
\item[i)]  For each $0 < a < I_{sup}(\gamma)$, the set $I(a)$ is a Meyer set.
\item[ii)] $\widehat{\gamma}_{pp}$ is a nontrivial sup almost periodic measure.
\end{itemize}
\end{theorem}

\section{A note on the dynamical spectra of a Meyer set}

In this section we will se an interesting consequence of Theorem \ref{aperMS} to the eigenfunctions of the dynamical system $\XX(\Lambda)$ corresponding to the set $\cup_{0 < \epsilon <1} \Delta^\epsilon \,$. Lets recall first the following Theorem.

\begin{theorem}\label{LSTHEOREM}\cite{LS} Let $m$ be a square integrable probability measure on the space of all measures on G, with associated autocorrelation\footnote{See \cite{BL} or \cite{LS} for the definition of the associated autocorrelation for $(\XX(\Lambda, m)$.}   $\gamma=\gamma_m$.
For $\varphi \in C_c (G)$ and $\lambda\in \hat{G}$, the following
  assertions are equivalent:

\begin{itemize}

\item[(i)] $|\widehat{\varphi}|^2(\lambda)\widehat{\gamma}(\{\lambda\}) >0$.

\item[(ii)] $E(\{\lambda\}) f_\varphi\neq 0$.

\item[(iii)] There exists an $f\neq 0$ with $f=E(\{\lambda\}) f$  in the closed convex  hull of $\{
  \overline{(\lambda,t)} T^t  f_\varphi : t\in G\}$.
\end{itemize}
\end{theorem}

Combining this result with Theorem \ref{aperMS}, we get:

\begin{cor}\label{eigenfun} Let $\Lambda$ be a Meyer set, $0< \epsilon <1$, $\chi \in \Delta^\epsilon$ and $m$ any ergodic measure on $\XX(\Lambda)$. Let $c \in C_c(\R^d)$ be so that $\widehat{c}(\chi) \neq 0$, and let $f_c : \XX(\Lambda) \rightarrow \C$ be defined by

$$f_c(\Lambda')= \sum_{x \in \Lambda'} c(-x) \,.$$

Then the closed convex  hull of $\{ \overline{(\chi,t)} T_t  f_c : t \in \R^d \}$ in $L^2(\XX(\Lambda), m)$ contains some eigenfunction $f_\chi$ corresponding to $\chi$.
\end{cor}
\noindent {\sc \bf Proof:}  We start by proving first the following Lemma:

\begin{lemma}\label{eiglem} Let $\Lambda$ be a set with Finite Local Complexity, let $\Delta=\Lambda-\Lambda$ and let $\Gamma \in \XX(\Lambda)$. Then
$$\Gamma-\Gamma \subset \Delta \,.$$
\end{lemma}
\noindent {\sc \bf Proof:} Let $x,y \in \Gamma$. Pick some $R>0$ so that $x,y \in \Gamma \cap B_R(0)$.

For each $n >0$ we can find some $t_n \in \R^d$ so that $\Gamma \cap B_n \subset T_{t_n}\Lambda +B_{\frac{1}{n}}$.

Thus, we can find some $x_n,y_n \in \Lambda$ so that $d(x, x_n-t_n) \leq \frac{1}{n}$ and $d(y, y_n-t_n)\leq \frac{1}{n}$.

Let $z_n =x_n-y_n \in \Delta$.

Then $d(x-y, z_n) \leq \frac{2}{n}$. This, shows that $x-y$ is in the closure of $\Delta$. But since $\Lambda$ has Finite Local Complexity, $\Delta$ is closed, thus

$$x-y \in \Delta \,.$$

\qed

We now return to the proof of Corollary \ref{eigenfun}.

Let $\gamma$ be the associated autocorrelation of $m$. Then, for $m$-almost all $\Gamma \in \XX(\Lambda)$, $\gamma$ is the autocorrelation of $\Gamma$ \cite{LS}.

Pick one such $\Gamma$. Then $ \Gamma - \Gamma \subset \Delta \,.$

In particular, $\Gamma$ is also a Meyer set.

Since $\chi \in \Delta^\epsilon \subset (\Gamma- \Gamma)^\epsilon$ and $0< \epsilon <1$, it follows from Theorem \ref{aperMS} that

$$\widehat{\gamma}(\{ \chi \}) \neq 0 \,.$$

Now our claim follows from Theorem \ref{LSTHEOREM}.

\qed

\section{Cordoba's Theorem on Crystals}\label{Cord}

One known result in crystallography, due to
Cordoba, says that if the Fourier transform of $\delta_\Gamma$
is pure point and $\Gamma$ is uniformly discrete, then we are in
the periodic crystal case. Since the original proof is very long, one would
like to get a simpler one. In this section we will see that, under the extra Meyer set assumption, we can prove stronger versions of this result.
Of course the Meyer assumption is a strong requirement, and thus our results are actually weaker than the Cordoba Theorem.

The Cordoba Theorem states:

\begin{theorem}{\rm \cite{CORD}} \label{S3COR} Suppose that the point sets $\Lambda_1,...,\Lambda_n$ are pairwise disjoint and
$\Gamma = \bigcup_{i=1}^n \Lambda_i$ is uniformly discrete. Let
$$\mu = \sum_{i=1}^n c_i \delta_{\Lambda_i} \,,$$
for some (different) complex numbers $c_1,...,c_n$. If
$\widehat{\mu}$ is a translation bounded pure point measure, then
each $\Lambda_i$ is a finite disjoint union of translates of
lattices.
\end{theorem}

The first result we get in this section, is an immediate consequence of Corollary \ref{cor123}.

\begin{theorem}\label{corsim1} Let $\mu$ be a positive translation bounded measure and let $\mu_{pp}=\sum_{x \in \Gamma} \mu(x)\delta_x$. Suppose that
$\mu$ is Fourier transformable,  $\widehat{\mu}$ is supported on
a Meyer set and that there exists an $a>0$ so that $\mu(x) > a$ for all $x \in \Gamma$.
Then $\Gamma$ is a subset of finitely many translates of a lattice. Moreover, $\supp(\widehat{\mu})$ is also a subset of finitely many translates of some lattice.
\end{theorem}
\noindent {\sc \bf Proof:} Since $\mu$ is positive, we get that $\widehat{\mu}$ is positive definite and thus Fourier Transformable. Moreover \cite{ARMA1}

$$\widehat{\widehat{\mu}}= \widetilde{\mu} \,.$$

Applying now Corollary \ref{cor123} to the measure $\widehat{\mu}$ we obtain the desired result.
\qed

If we compare Theorem \ref{corsim1} to Theorem \ref{S3COR}, we don't require that $\mu$ is a discrete measure, and we also weakened the requirement that $\mu$ has uniformly discrete support and only takes finitely many values to asking that $\mu(\{x\})$ doesn't come arbitrarily close to $0$. But we added two conditions: $\mu$ positive and $\supp(\widehat{\mu})$ is a Meyer set.

We conclude the section by providing a very simple proof of Theorem \ref{S3COR}, under the extra assumption $\supp(\mu)$ is a Meyer set:

\begin{prop}\label{corsim2}
Suppose that the point sets $\Lambda_1,...,\Lambda_n$ are pairwise
disjoint and $\Lambda = \bigcup_{i=1}^n \Lambda_i$ is a Meyer set.
Let
$$\mu = \sum_{i=1}^n c_i \delta_{\Lambda_i} \,,$$ for some pairwise distinct nonzero complex numbers $c_1,...,c_n$.

If $\widehat{\mu}$ is a discrete Fourier Transformable measure,
then each $\Lambda_i$ is a finite union of translates of the same
lattice.
\end{prop}

\noindent {\sc \bf Proof:}  Let $K$ be a compact set with non-empty interior, so that $(\Lambda-\Lambda) \cap K = \{ 0 \}$. Such a set exists because
$\Lambda-\Lambda$ is uniformly discrete.

Recall that on the space of translation bounded measure we can define a norm $\| \, \|_K$ by

$$\| \nu \|_K = \sup_{x \in \R^d} \left| \nu \right| (x+K) \,.$$

A measure $\nu$ is called norm almost periodic, if for each $\epsilon >0$ the set

$$P_\epsilon(\nu) = \{ t \in \R^d | \| \nu -T_t \nu \|_K < \epsilon \} \,,$$

is relatively dense.

Since $\widehat{\mu}$ is pure point diffractive, and $\supp(\mu)$ is Meyer, it follows that $\mu$ is norm almost periodic \cite{BM}.

Now we prove that $\mu$ is fully periodic. Pick some
$$0< \epsilon < \min \{ \left| c_i-c_j \right| | i\neq j \} \cup \{ \left|c_i \right| | 1 \leq i \leq n\} \,.$$

We will show that any $\epsilon$-norm almost period of $\mu$ is also a period.

Let $t \in P_\epsilon(\mu)$ and $x \in \R^d$. Then

$$\left| \mu( \{x \}) - \mu(\{ x+t \} ) \right| < \epsilon \,.$$

We know that $\mu( \{x \}) , \mu(\{ x+t \} ) \in A:= \{ c_i | 1 \leq i \leq n \} \cup \{0 \}$.

Also, by the definition of $\epsilon$, if $a,b \in A$ with $a \neq b$, we have $\left| a -b \right| > \epsilon$. Hence

$$\mu( \{x \}) = \mu(\{ x+t \} ) \,.$$

Thus, since $x \in \R^d$ is arbitrary, we get that $t$ is a period for $\mu$.

Let
$$L:= \{ t \in \R^d | T_t \mu = \mu \} \,,$$

be the group of periods of $\mu$. Then, $P_\epsilon(\mu) \subset L$, which shows that $L$ is relatively dense.
Also, since $\supp(\mu) \subset \Lambda$, it is easy to show that $L \subset \Lambda- \Lambda$, hence $L$ is also uniformly discrete.

Thus $L$ is a lattice.

Now the rest of the proof is simple. Since $c_i$ are nonzero and pairwise distinct, it follows immediately that $\Lambda_i +L = \Lambda_i$.

Let $K_0$ be a fundamental domain for $L$, and let $F_i := \Lambda_i \cap K_0$. Then

$$F_i+L \subset \Lambda_i +L = \Lambda_i \,.$$

Also, any $x \in \Lambda_i$ can be written as $x=y+z$ with $y \in L$ and $z \in K_0$. But then

$$z=x-y \in \Lambda_i +L = \Lambda_i \,.$$

Thus, $z \in \Lambda_i \cap K_0 =F_i$. This shows that $x \in L+ F_i$, and hence

$$F_i+ L \subset \Lambda_i \subset F_i+L \,.$$
\qed

\section{Small Deformations of Meyer sets }\label{S3swlms}

Given a Meyer set $\Lambda$, we know that the set $\mB$ of Bragg peaks is relatively dense. But what happens if we deform $\Lambda$?
In this section we show that if the deformation is small in density, then the new point set still shows a relatively dense set of Bragg peaks.

Our approach is simple: We first show that if $\eta, \eta_1$ are Fourier Transformable measures, with $\eta-\eta_1$ positive, then $\left| \widehat{\eta}(\{ \chi \})-  \widehat{\eta_1}(\{ \chi \})\right|$ attains its maximum at $\chi=0$.

From here it follows that if $\Lambda \subset \Omega$, then $\left| \widehat{\gamma_\Lambda}(\{ \chi \})-  \widehat{\eta_\Omega}(\{ \chi \})\right|$
is bounded by $ \widehat{\gamma_\Lambda}(\{ 0 \})-  \widehat{\eta_\Omega}(\{ 0 \})$, a difference which can be related to the densities of the two point sets. Thus, we will get a simple density bound for the difference $\left| \widehat{\gamma_\Lambda}(\{ \chi \})-  \widehat{\eta_\Omega}(\{ \chi \})\right|$; which implies that as long as one of $\widehat{\gamma_\Lambda}(\{ \chi \}),  \widehat{\eta_\Omega}(\{ \chi \})$ exceeds this bound, the other is non-zero.

Finally, if $\Gamma$ is a deformation of $\Lambda$, then both $N:= \Gamma \cap \Lambda$ is a subset of both $\Lambda$ and $\Gamma$, and thus the above considerations allow us to go from $\Lambda$ first to $N$ and then to $\Gamma$.

\begin{lemma}\label{L2112}
Let $\eta, \eta_1$ be Fourier transformable measures with
$\eta-\eta_1$ positive. Then, for all $\chi \in  \R^d$ we have:
\begin{equation}
\left| \widehat{\eta}(\{ \chi \})-  \widehat{\eta_1}(\{ \chi \})
\right| \leq \widehat{\eta}(\{ 0 \})-  \widehat{\eta_1}(\{ 0 \})
\,.
\end{equation}
\end{lemma}

\noindent {\sc \bf Proof:}

This result follows immediately from Theorem \ref{HOFave}:
\begin{equation*}
\begin{split}
\left| \widehat{\eta}(\{ \chi \})-  \widehat{\eta_1}(\{ \chi \})
\right| &= \lim_n \frac{ \left| \int_{A_n} \chi(x) d \eta(x) -
\int_{A_n} \chi(x) d \eta_1(x) \right|}{\Vol(A_n)} \\
&\leq \lim_n \frac{ \int_{A_n} |\chi(x) | d \left| \eta- \eta_1\right|(x)
}{\Vol(A_n)}= \lim_n \frac{ \int_{A_n} 1 d  (\eta- \eta_1)(x)
}{\Vol(A_n)} \\
&= \widehat{\eta}(\{ 0 \})-  \widehat{\eta_1}(\{ 0 \}) \,.
\end{split}
\end{equation*}
\qed

An immediate consequence of this Lemma is the following result:

\begin{prop}\label{slms} Let $\omega_1, \omega_2$ be two translation bounded measures, with autocorrelations $\gamma_1, \gamma_2$. Suppose that
$$\gamma_1-\gamma_2 \geq 0 \,.$$

Then, for all $\chi \in \R^d$ we have

\begin{eqnarray*}
\widehat{\gamma_1}(\{ \chi \}) \geq  \widehat{\gamma_2}(\{ \chi \}) + \widehat{\gamma_2}(\{ 0 \})
- \widehat{\gamma_1}(\{ 0 \}) \,, \end{eqnarray*} and
 \begin{eqnarray*}\widehat{\gamma_2}(\{ \chi \}) \geq  \widehat{\gamma_1}(\{ \chi \}) + \widehat{\gamma_2}(\{ 0 \})
- \widehat{\gamma_1}(\{ 0 \}) \,.\end{eqnarray*}
\end{prop}

If $0 \leq \omega_1 \leq \omega_2$, then it is easy to show that $0 \leq \gamma_1 \leq \gamma_2$. For this reason, $0 \leq \omega_1 \leq \omega_2$ will be a standard assumption we will make in most of the results, since we will often need the condition $\gamma_2-\gamma_1 \geq 0$.
Note that for $\omega_1= \delta_\Lambda, \omega_2= \delta_\Gamma$, the condition $0 \leq \omega_1 \leq \omega_2$ is equivalent to $\Lambda \subset \Gamma$.

Thus, if $\Lambda \subset \Gamma$, Proposition \ref{slms} yields:

\begin{cor}\label{S3h1}
Let $\Lambda \subset \Gamma$ and let $\gamma_\Lambda, \gamma_\Gamma$ denote
their autocorrelation measures, and let $a >
\widehat{\gamma_\Gamma}(\{ 0 \}) - \widehat{\eta_\Lambda}(\{ 0 \})$.
\begin{itemize}
\item[i)] If $\Lambda$ has a relatively dense set of $a$-visible
Bragg peaks, then $\Gamma$ has a relatively dense set of Bragg peaks.
\item[ii)] If $\Gamma$ has a relatively dense set of $a$-visible Bragg
peaks, then $\Lambda$ has a relatively dense set of Bragg peaks.
\end{itemize}
\end{cor}

If $\Lambda$ is a Meyer set, then we know that for all $a < \widehat{\gamma_\Lambda}(\{0\})$ the set of
$a$-visible Bragg peaks is relatively dense. For the remaining of the section, we will try to combine this result with Corollary \ref{S3h1}, and then replace the difference  $\widehat{\gamma_\Gamma}(\{ 0 \}) - \widehat{\eta_\Lambda}(\{ 0 \})$ by an expression involving the densities of $\Lambda$ and $\Gamma$.

\begin{prop} \label{4.742} Let $\Lambda$ a Meyer set, $\Delta : \Lambda- \Lambda$ and let $\Gamma$ be a Delone set.
Let $N:= \Lambda \cap \Gamma$, and let $\gamma_N, \gamma_\Gamma$ denote autocorrelations of $N$ respectively $\Gamma$.  If
$$2\widehat{\gamma_N}(\{ 0 \}) > \widehat{\gamma_\Gamma}(\{ 0 \})\,,$$
then there exists an $\epsilon >0$ so that $\Delta^\epsilon$ is a
subset of the Bragg spectrum of $\Gamma$. In particular,
$\Gamma$ has a relatively dense set of Bragg peaks.
\end{prop}
\noindent {\sc \bf Proof:}

It follows from Lemma \ref{L2112} that for all $\chi \in \R^d$ we have
$$\widehat{\gamma_\Gamma} (\{ \chi \}) \geq \widehat{\gamma_N} (\{
 \chi \}) + \widehat{\gamma_N} (\{
0\})- \widehat{\gamma_\Gamma} (\{ 0\}) \,.$$

Also, since $N-N \subset \Delta$, we get from Theorem \ref{S31} that there exists a constant $C>0$ so that, for all $\chi \in \Delta^\epsilon$ we have

$$ \left| \widehat{\gamma_N} (\{
 \chi \}) - \widehat{\gamma_N} (\{ 0\}) \right| < C\epsilon \,.$$

In particular, for all $\chi \in \Delta^\epsilon$ we have

$$ \widehat{\gamma_N} (\{ \chi \}) \geq \widehat{\gamma_N} (\{ 0\})   - C\epsilon \,,$$

and thus

$$\widehat{\gamma_\Gamma} (\{ \chi \}) \geq 2\widehat{\gamma_N} (\{
0\})- \widehat{\gamma_\Gamma} (\{ 0\}) -C\epsilon \,.$$

The rest of the proof is now clear. Pick some
$$0< \epsilon < \frac{2\widehat{\gamma_N} (\{ 0\})- \widehat{\gamma_\Gamma} (\{ 0\})}{C} \,.$$ Then, for all $\chi \in \Delta^\epsilon$ we have

$$\widehat{\gamma_\Gamma} (\{ \chi \}) \geq 2\widehat{\gamma_N} (\{
0\})- \widehat{\gamma_\Gamma} (\{ 0\}) -C\epsilon >0 \,.$$

 \qed

For the rest of the section we try to replace the condition
$2\widehat{\gamma_N}(\{ 0 \}) - \widehat{\gamma_\Gamma}(\{ 0 \})> 0$ by
one that is easier to understand.

If $\Gamma$ is uniformly distributed, then it is known \cite{HOF}, \cite{LR} that

$$\widehat{\gamma_{\Gamma}}( \{ 0 \})= \dens(\Gamma)^2 \,.$$

This would allow us to replace  $\widehat{\gamma_N}(\{ 0 \})$ and $\widehat{\gamma_\Gamma}(\{ 0 \})$ by the densities of the two sets, but we would need both sets to be uniformly distributed, a very strong requirement. Instead, we will prove that for arbitrary Delone sets, the above formula can be replaced by a inequalities involving the lower and upper density of the set. This will allow us weaken the restrictions on $N$ and $\Gamma$

We introduce now the concept of lower and upper density:

\begin{defi} \rm For a Delone set $\Gamma \subset \R^d$ and a
fixed van Hove sequence $\{ B_n \}$, we define the {\bf lower and
upper density} of $\Gamma$ by:
$$\underline{{\rm dens}(\Gamma)} := \liminf_{n \to \infty} \inf_{x
\in \R^d} \frac{\sharp ( \Gamma \cap (x +B_n))}{ \Vol (B_n)} \,,$$

$$\overline{{\rm dens}(\Gamma)} := \limsup_{n \to \infty} \sup_{x
\in \R^d} \frac{\sharp ( \Gamma \cap (x +B_n))}{ \Vol (B_n)
} \,.$$

A point set $\Gamma$ is called {\bf uniformly distributed} if
$$\underline{{\rm dens}(\Gamma)} =\overline{{\rm dens}(\Gamma)}
=: {\rm dens}(\Gamma) \,.$$
\end{defi}

We now prove the following Lemma:

\begin{lemma}\label{lll222} Let $\Gamma$ be a Delone set and assume that its
autocorrelation $\gamma$ exists for our van Hove sequence $\{ B_n
\}$. Then
$$ ( \underline{{\rm dens}(\Gamma)} )^2 \leq
\widehat{\gamma}(\{0\}) \leq (\overline{{\rm dens}(\Gamma)})^2
\,.$$
\end{lemma}
\noindent {\sc \bf Proof:} Let $\epsilon >0$ be fixed but
arbitrarily. Since
\begin{equation*}
\widehat{\gamma}(\{0\}) = \lim_{n \to \infty} \frac{
\gamma(-B_n)}{\Vol (B_n) } \,,
\end{equation*} then, there exists some $n_0>0$ so that for all $n>n_0$ and all $x
\in \R^d$ we have

\begin{equation}\label{ugly1}
\left| \widehat{\gamma}(\{0\}) - \frac{ \gamma(-B_n)}{\Vol (B_n)
} \right| < \epsilon \,,
\end{equation}

\begin{equation*}
 \underline{{\rm dens}(\Gamma)}  < \frac{ \sharp(\Gamma \cap (x+B_n))}{\Vol (B_n)
} + \epsilon \,,
\end{equation*}

and

\begin{equation*}
 \overline{{\rm dens}(\Gamma)}  > \frac{ \sharp(\Gamma \cap (x+B_n))}{\Vol (B_n)
} - \epsilon \,.
\end{equation*}

By \cite{BL} we have
$$\gamma =\lim_{m \to \infty} \frac{ \delta_{\Gamma \cap B_m} *
\widetilde{\delta_{\Gamma}}}{\Vol (B_m) } \,.$$

Thus  there exists an $m_0$ so that, for all $m >m_0$ we have

\begin{equation}\label{ugly4}
\left|  \frac{ \gamma(-B_n)}{\Vol (B_n)}- \frac{
\delta_{\Gamma \cap B_m} * \widetilde{\delta_{\Gamma}}(-B_n)}{\Vol (B_m) \Vol (B_n) } \right| < \epsilon \,.
\end{equation}

\noindent(\ref{ugly1}) and (\ref{ugly4}) allow us relate $\widehat{\gamma}(\{0\})$ to $\delta_\Gamma$. By combining these two relations,
for all $m > \max \{ m_0, n_0 \}$ we have

\begin{equation}\label{ugly5}
\left| \widehat{\gamma}(\{0\}) - \frac{ \delta_{\Gamma \cap B_m} *
\widetilde{\delta_{\Gamma}}(-B_n)}{\Vol (B_m) \Vol (B_n) }  \right| < 2\epsilon \,.
\end{equation}

A simple computation shows that

\begin{equation}\label{uglyfication}
\begin{split}
\delta_{\Gamma \cap B_m} * \widetilde{\delta_{\Gamma}}(-B_n) &=
\sharp \{ (x,y) \mid x \in \Gamma \cap B_m , y \in \Gamma , x -
y \in -B_n
\} \\
&= \sharp \{ (x,y) \mid x \in \Gamma \cap B_m , y \in \Gamma
\cap
(x+B_n) \} \\
&= \sum_{ x \in \Gamma \cap B_m} \sharp (\Gamma \cap (x+B_n))
\,.
\end{split}
\end{equation}

which allows us to relate $\delta_{\Gamma \cap B_m} * \widetilde{\delta_{\Gamma}}(-B_n)$ to the lower and upper density of $\Gamma$. Indeed, since

\begin{equation*}
 \underline{{\rm dens}(\Gamma)} - \epsilon  < \frac{ \sharp(\Gamma \cap (x+B_n))}{\Vol (B_n)} ;  \frac{ \sharp(\Gamma \cap (B_m))}{\Vol (B_m)} <  \overline{{\rm dens}(\Gamma)} + \epsilon  \,,
\end{equation*}

by (\ref{uglyfication}) we get

$$(\underline{{\rm dens}(\Gamma)} - \epsilon)^2  < \frac{\delta_{\Gamma \cap B_m} * \widetilde{\delta_{\Gamma}}(-B_n)}{\Vol(B_n) \Vol(B_m) } <  (\overline{{\rm dens}(\Gamma)} + \epsilon)^2  \,.$$

Now the desired result follows immediately from (\ref{ugly5}):

$$
\widehat{\gamma}(\{0\}) \geq  \frac{ \delta_{\Gamma \cap B_m} *
\widetilde{\delta_{\Gamma}}}{\Vol (B_m )\Vol ( B_n)} (-B_n) - 2\epsilon \geq (\underline{{\rm dens}(\Gamma)} - \epsilon)^2 -
 2\epsilon \,,
$$

 and

$$
\widehat{\gamma}(\{0\}) \leq  \frac{ \delta_{\Gamma \cap B_m} *
\widetilde{\delta_{\Gamma}}}{\Vol (B_m )\Vol ( B_n)} (-B_n) + 2\epsilon \leq (\overline{{\rm dens}(\Gamma)} + \epsilon)^2 + 2\epsilon \,.
$$

Thus , for all $\epsilon >0$ we have

$$(\underline{{\rm dens}(\Gamma)} - \epsilon)^2 -
 2\epsilon \leq \widehat{\gamma}(\{0\}) \leq (\overline{{\rm dens}(\Gamma)} + \epsilon)^2 +
 2\epsilon \,,$$

 which completes the proof.
\qed

Suppose now that $\Gamma$ is a Delone set, and $\Lambda$ is Meyer. Let $N:= \Lambda \cap \Gamma$.  Proposition \ref{4.742}
tells us that the condition

$$2\widehat{\gamma_N}(\{ 0 \}) > \widehat{\gamma_\Gamma}(\{ 0 \})\,,$$

is enough to guarantee that $\Gamma$ has a relatively dense set of Bragg peaks, while Lemma \ref{lll222} allows us replace the intensities of the Bragg peaks at $0$ by the lower and upper density of $N$. Thus, we get:

\begin{cor}\label{finalj}  Let $\Lambda$ a Meyer set, $\Delta = \Lambda- \Lambda$, let $\Gamma$ be a Delone set and let $N=\Lambda \cap \Gamma$ . If

$$\sqrt{2}\underline{\rm dens}(N)  > \sqrt{2} \overline{\rm dens}(\Gamma) \,,$$

then there exists an $\epsilon >0$ so that $\Delta^\epsilon$ is a
subset of the Bragg spectrum of $\Gamma$.
\end{cor}

We will finish this section by obtaining an upper bound for $\overline{\rm dens}(\Lambda \bigtriangleup \Gamma)$ in terms of upper and lower density of $\Lambda$, which implies the condition in Corollary \ref{finalj}.

Since $\Gamma= \Lambda \bigtriangleup (\Lambda \bigtriangleup \Gamma)$, $\Gamma$ is a $\Lambda \bigtriangleup \Gamma$ deformation of $\Lambda$. Theorem \ref{dens} below states that if the deformation is small in density compared to $\Lambda$, then $\Gamma$ keeps some of the Bragg spectra of $\Lambda$.

\begin{theorem}\label{dens} Let $\Lambda$ a Meyer set, $\Delta : \Lambda- \Lambda$, and let $\Gamma$ be a Delone set. If

$$\overline{\rm dens}(\Lambda \bigtriangleup \Gamma) < \frac{\sqrt{2}\underline{\rm dens}(\Lambda)-\overline{\rm dens}(\Lambda)}{\sqrt{2}+1}  \,,$$

then there exists an $\epsilon >0$ so that $\Delta^\epsilon$ is a
subset of the Bragg spectrum of $\Gamma$.

\end{theorem}
\noindent {\sc \bf Proof:} We prove first that

\begin{equation}\label{junkfinal1}
\underline{\rm dens}(\Lambda \cap \Gamma)+ \overline{\rm dens}(\Lambda \backslash \Gamma) \geq  \underline{\rm dens}(\Lambda)
\end{equation}

Let $\epsilon >0$.

By the definition of $\underline{\rm dens}(\Lambda)$, there exists an $n_0$ so that, for all $n>n_0$ and for all $x$ we have

$$\frac{\sharp ( \Lambda \cap (x +B_n))}{ \Vol (B_n)} > \underline{\rm dens}(\Lambda)  - \epsilon \,.$$

Thus, for all $n > n_0$ we have

$$\frac{\sharp \left[ (\Lambda \cap \Gamma) \cap (x +B_n)\right]}{ \Vol (B_n)} +\frac{\sharp \left[ (\Lambda \backslash \Gamma) \cap (x +B_n)\right]}{ \Vol (B_n)}  > \underline{\rm dens}(\Lambda)  - \epsilon \,.$$

Using now the definition of $\overline{\rm dens}(\Lambda \backslash \Gamma)  $, we get an $n_1$ so that, for all $n > n_1$, and all $x$ we have

$$\frac{\sharp \left[ (\Lambda \backslash \Gamma) \cap (x +B_n)\right]}{ \Vol (B_n)}  < \overline{\rm dens}(\Lambda \backslash \Gamma)  + \epsilon $$

Thus, for all $n > \max\{ n_0, n_1 \}$ we have

$$\frac{\sharp \left[ (\Lambda \cap \Gamma) \cap (x +B_n)\right]}{ \Vol (B_n)}  \geq  \underline{\rm dens}(\Lambda)  -\overline{\rm dens}(\Lambda \backslash \Gamma) -2\epsilon \,.$$

From which (\ref{junkfinal1}) follows immediately.

Let $N:= \Gamma \cap \Lambda$. We will show next that (\ref{junkfinal1}) together with the condition from the Theorem imply

$$\sqrt{2}\underline{\rm dens}(N)  > \sqrt{2} \overline{\rm dens}(\Gamma) \,.$$

Indeed, by (\ref{junkfinal1}) we have

$$\underline{\rm dens}(N)\geq  \underline{\rm dens}(\Lambda) - \overline{\rm dens}(\Lambda \backslash \Gamma) \,.$$

Also, the relation in the Theorem can be rewritten as

$$\sqrt{2}\underline{\rm dens}(\Lambda) > (\sqrt{2}+1)\overline{\rm dens}(\Lambda \bigtriangleup \Gamma)+\overline{\rm dens}(\Lambda) \,.$$

Hence,

\begin{eqnarray*}
\begin{split}
\sqrt{2}\underline{\rm dens}(N) &\geq  \sqrt{2}\underline{\rm dens}(\Lambda) - \sqrt{2}\overline{\rm dens}(\Lambda \backslash \Gamma)\\
&\geq (\sqrt{2}+1)\overline{\rm dens}(\Lambda \bigtriangleup \Gamma)+\overline{\rm dens}(\Lambda)  - \sqrt{2}\overline{\rm dens}(\Lambda \backslash \Gamma) \\
&\geq \overline{\rm dens}(\Lambda \bigtriangleup \Gamma)+\overline{\rm dens}(\Lambda) \\
&\geq \overline{\rm dens}(\Gamma \backslash \Lambda)+\overline{\rm dens}(\Gamma \cap \Lambda) \geq \overline{\rm dens}(\Gamma) \,.
\end{split}
\end{eqnarray*}

which completes the proof.

\medskip

\subsection*{Acknowledgments} I wish to thank Dr. Robert Moody, Dr. Michael Baake and Dr. Daniel Lenz for their support, guidance, useful
suggestions during the preparation of this paper. Part of this work was supported by the Natural Sciences
and Engineering Research Council of Canada.

\printindex
\newpage
\appendix
\addcontentsline{toc}{chapter}{Appendices}
\section{Strong Meyer sets}\label{S2appen}

In this section we try to generalize the results from this paper
to the case of locally compact Abelian groups. The main problem we have to face is that the proof of Theorem \ref{S3CMS} relies on the geometry of $\R^d$. Thus, it is not known in general what relationship, if any, is among those conditions.

Thus, it is not clear how one should define a Meyer set in an arbitrary locally compact Abelian group. In the
spirit of the paper, we will use the definition based on the relative
denseness of the $\epsilon$-dual characters to define a strong Meyer set.

For the entire section $G$ represents a  $\sigma-$compact, locally compact Abelian
group. $\theta$ will denote its Haar measure, while $\{ A_n \}_n$ will be a fixed van Hove sequence.

We start by reviewing some results from \cite{ARMA} which we will need in this Section. We start by seeing that in general it is possible to go from $\widehat{\mu}$ back to $\mu$:

\begin{prop}{\rm \cite{ARMA1}}
Let $\mu$ be a transformable measure on $G$, with $\widehat{\mu}$
transformable. Then
$$\widehat{\widehat {\mu}}=\widetilde{\mu}\,.$$
\end{prop}

This result will play an important role for the rest of the Section, since the results in \cite{ARMA} are usually the dual results to the ones we need. Thus, we will generally need to apply their result to $\widehat{\mu}$. This is the reason why many of the results below require that $\mu$ is twice Fourier Transformable.

\begin{prop}\label{mea} If $\mu \in {\mathcal M}^\infty(G)$ is Fourier transformable with $\widehat{\mu}$
Fourier transformable, then for any $\psi \in \widehat{G}$ we
have:

$$\widehat{\mu}(\{ \psi \})= \lim_{n \to \infty} \frac { \int_{A_n} \psi(-x) d \mu(x) }{\theta (A_n ) }\,.$$
\end{prop}

\noindent {\sc \bf Proof:} Since $\widehat{\mu}$ is Fourier
transformable, by applying the Theorem 11.3 from \cite {ARMA} to
the inverse Fourier transform of $\mu$ we get:
$$\widehat{\mu}(\{ \psi \})= M(\psi^{-1}\mu )=  \lim_{n \to \infty} \frac { \int_{A_n} \psi(-x) d \mu(x) }{\theta (A_n ) } \,.$$
 \qed

\begin{cor} If $\mu \in{\mathcal M}^\infty(G)$ is Fourier transformable and positive, then for any $\psi \in
\widehat{G}$ we have:
$$\left| \widehat{\mu}(\{ \psi \}) \right| \leq  \widehat{\mu}(\{ 0 \}) \,.$$
\end{cor}

\noindent {\sc \bf Proof:} Since $\mu$ is positive, $\widehat{\mu}$ is a positive definite measure, and thus Fourier Transformable \cite{BF}.

Thus, by using Proposition \ref{mea} we get:

$$\left|  \widehat{\mu}(\{ \psi \})\right| = \lim_{n \to \infty} \left |
\frac {\int_{A_n} \psi(-x) d \mu(x) }{\theta (A_n ) }\right | \leq
\lim_{n \to \infty}  \frac { \int_{A_n} \left| \psi(-x)\right| d
\mu(x)}{\theta (A_n ) } = \widehat{\mu}(\{ 0 \}) \,.$$ \qed

We can now prove a result similar to Theorem \ref{S31}:

\begin{prop} \label{111} Let $\mu \in {\mathcal M}^\infty (G)$ be a positive Fourier transformable measure. Let $\Delta$ be any set such that
${\rm supp}(\mu)\subset \Delta$. Let $\epsilon >0$ be arbitrary. Then, for all $\psi \in
\Delta^\epsilon$ and $\chi \in \widehat{G}$ we have:
$$\left| \widehat{\mu }(\{ \psi+ \chi\})-\widehat{\mu }(\{\chi\}) \right| \leq \epsilon \widehat{\mu}(\{ 0 \})\,.$$

\end{prop}

\noindent {\sc \bf Proof:}
\noindent

\noindent Let $\psi \in \Delta^{\epsilon}$ and $\chi \in
\widehat{G}$. Then, for all  $x \in \Delta$ we have:

$$\left|  (\psi+\chi)(-x)  - \chi(-x)  \right| =\left|  \overline{\psi}(x)  - 1  \right| < \epsilon \,.$$

\noindent Using the fact that $\supp(\gamma) \subset \Delta$ we get:

\begin{equation*}
\begin{split}
\left| \widehat{\mu}(\{ \psi+\chi\})-\widehat{\mu}(\{\chi\})
\right| &\leq \lim_{n \to \infty} \frac { \int_{A_n} \left|
(\psi+\chi)(-x) -(\chi)(-x)\right|\, d \left| \mu \right|(x)
}{\theta (A_n ) } \\
&\leq \lim_{n \to \infty} \frac { \int_{A_n} (\epsilon) \, d
\left| \mu \right|(x) }{\theta (A_n ) } = \epsilon \lim_{n \to \infty}
 \frac { \left| \mu \right|(A_n) }{\theta (A_n ) } \\
&=\lim_{n \to \infty}\epsilon \frac { \mu (A_n) }{\theta (A_n ) }= \epsilon \widehat{\mu}(\{ 0 \})
\,.
\end{split}
\end{equation*}

\qed

We can now introduce the notion of strong Meyer set and look at its diffraction.

\begin{defi} \rm Let $\Lambda \subset G$ and let $\Delta := \Lambda-\Lambda$.
We say that $\Lambda$ is a {\bf strong Meyer set} if:
\begin{itemize}
\item[i)] $\Lambda$ is a Delone set with finite local complexity,
\item[ii)] $\Delta^\epsilon$ is relatively dense in $\widehat{G}$ for all $\epsilon >0$.
\end{itemize}
\end{defi}

The condition $\Lambda$ has finite local complexity is needed to make sure that any autocorrelation $\gamma$ of $\Lambda$ is supported inside $\Delta$.

It is easy to see that in $\R^d$ a set $\Lambda$ is a strong Meyer set if and only if
$\Lambda$ is a Meyer set.

For the rest of the section, $\Lambda$ is a strong Meyer set. As usual we set $\Delta= \Lambda- \Lambda$ and
$\gamma$ is an autocorrelation  of $\Lambda$. $\gamma$ is a positive and positive
definite measure, thus twice Fourier transformable.

We use again $I(a)$ to denote the set of $a$-visible Bragg peaks, that is:
$$I(a): =\{ \chi \in \widehat{G} \mid \widehat{ \gamma } (\{ \chi \}) \geq a \} \,.$$

The following two results can be proved like in Section \ref{S3TBMWMSS}, and we skip their proofs.

\begin{prop} Let $\Lambda \subset G$ be a strong Meyer set, $\Delta = \Lambda - \Lambda$ and $I(a)$ denote the set of $a$-visible Bragg peaks. Then
\begin{itemize}
\item[i)] For all $\epsilon >0$ we have:
$$ I(a) \pm \Delta^{\epsilon} \subset I(a - \epsilon\widehat{\gamma}(\{ 0\})) \,.$$
\item[ii)] If $\widehat{\gamma}(\{ 0 \}) \neq 0$, then for all $0 < a < \widehat{\gamma}(\{ 0 \})$ there exists an
$\epsilon >0$ and a finite set $F$ such that:
$$ \Delta^{\epsilon} \subset I(a) \subset \Delta^{\epsilon} +F \,.$$
In particular $I(a)$ is relatively dense.
\end{itemize}
\end{prop}

\begin{prop}
Suppose that the set of Bragg peaks of a strong Meyer set $\Lambda$ is nontrivial and equal to $I(a)$ for some $a>0$. Then there exists a lattice
$L$, with dual lattice $L^{\ast}$, such that $\Lambda$ is a subset
of finitely many translates of $L$ and the set of Bragg peaks is a
subset of finitely many translates of $L^{\ast}$.
\end{prop}

\section{Vague Topology} \label{A2}

Given a Meyer set $\Lambda$, we had seen in Section \ref{S3TBMWMSS} that the $\epsilon$ dual characters $\Delta^\epsilon$ are sup almost periods
for the discrete part of the spectra $\widehat{\gamma}_{pp}$ of $\Lambda$.

An interesting Question is if there is any connection between $\Delta^\epsilon$ and the continuous spectrum $\widehat{\gamma}_{c}$.
When studying this measure, the sup topology is useless, so we need to look at a different topology.

In this Section we will show that, for a Meyer set $\Lambda \subset \R^d$, with autocorrelation $\gamma$, the measures $\widehat{\gamma}, \widehat{\gamma}_{pp}$ and $\widehat{\gamma}_{c}$ are almost periodic in the vague topology, and any set of almost periods of these measures contains some $\epsilon$ dual characters of $\Lambda$. All the results of this section will follow from the continuity of the Fourier Transform with respect to the vague topology.

First recall that since $\gamma$ is positive definite, it is weakly almost periodic, thus it can be written in an unique way as

$$\gamma= \gamma_S+\gamma_0 \,,$$

where $\gamma_S$ is a strong almost periodic measure and $\gamma_0$ is null weakly almost periodic \cite{ARMA}.

Moreover, since $\gamma$ is also positive, it follows that $\widehat{\gamma}$ is positive definite, thus both $\gamma$ and $\widehat{\gamma}$ are Fourier Transformable. Thus, by applying Theorem 11.2 in \cite{ARMA} to $\widehat{\gamma}$ we get:

$$ (\widehat{\gamma})_{pp}= (\widehat{\gamma_S})\,,$$
$$ (\widehat{\gamma})_{c}= (\widehat{\gamma_0})\,.$$

We also need the following result from \cite{NS}:

\begin{lemma}\label{L1NS}{\rm \cite{NS}} Let $\Lambda'$ be any regular model set containing $\Lambda$ and let $\Gamma:= \Lambda' - \Lambda'$. Then
\begin{equation*}
\begin{split}
{\rm supp}(\gamma) &\subset \Gamma \,, \\
{\rm supp}(\gamma_S) &\subset \Gamma \,, \\
{\rm supp}(\gamma_0) &\subset \Gamma \,.
\end{split}
\end{equation*}
\end{lemma}

Now, we can prove that the sets $\Gamma^\epsilon$ are sets of vague almost periods of $\widehat{\gamma}$, $\widehat{\gamma}_{pp}$ and $\widehat{\gamma}_{c}$.

\begin{prop} \label{ghj}
Let $U$ be any neighborhood of $0$ in the vague topology. Then,
there exists an $\epsilon > 0$ so that, for all $\chi \in
\Gamma^\epsilon$ we have:
\begin{itemize}
\item[i)] $\widehat{\gamma} - T_\chi (\widehat{\gamma}) \in U$,
\item[ii)] $(\widehat{\gamma})_{pp} - T_\chi ((\widehat{\gamma})_{pp})
\in U$, \item[iii)] $(\widehat{\gamma})_{c} - T_\chi
((\widehat{\gamma})_{c}) \in U$.
\end{itemize}
\end{prop}
\noindent {\sc \bf Proof:} Since the Fourier transform is
continuous in the vague topology, there exists $V$ an open
neighborhood of $0$ such that, for all $\mu \in V$ we have
$\widehat{\mu} \in U$.

Since the norm topology is stronger than the vague topology \cite{BM}, there exists a $\delta >0$ so that

$$\| \nu \|_K < \delta \Rightarrow \nu \in V \,.$$

Pick an $\epsilon >0$ so that

$$ \epsilon \| \gamma \|_K < \delta \,;\,  \epsilon \| \gamma_S \|_K < \delta \,{\rm and} \, \epsilon \| \gamma_0 \|_K < \delta \,.$$

We show now that for all $\chi \in \Gamma^\epsilon$ we have

$$\chi \gamma- \gamma \,;\, \chi \gamma_S - \gamma_S \,;\, \chi \gamma_0 - \gamma_0 \in V \,.$$

Indeed

\begin{eqnarray*}
\begin{split}
\| \chi \gamma - \gamma \|_K &= \sup_{x \in \R^d} \left| \chi \gamma - \gamma \right| (x+K)  =\sup_{x \in \R^d} \left| \int_{x+K} \chi(t)-1 d \gamma(t) \right| \\
&\leq  \sup_{x \in \R^d} \sum_{t \in (x+K) \cap \Gamma} \left|  \chi(t)-1 \right| \cdot \left| \gamma(\{ t \}) \right| \\
 &\leq \sup_{x \in \R^d} \epsilon \left| \gamma \right|(x+K) = \epsilon \| \gamma \|_K <  \delta \,,
\end{split}
\end{eqnarray*}

and thus

$$\chi \gamma - \gamma \in V \,.$$

Exactly the same way we can prove

$$\chi \gamma_S - \gamma_S, \chi \gamma_0 - \gamma_0 \in V \,.$$

By applying the Fourier Transform, and using $\widehat{\chi \mu}= T_\chi \widehat{\mu}$ we get the desired result.

\qed

\end{document}